\documentclass[aps,pra,reprint,superscriptaddress,nobibnotes]{revtex4-1}
\usepackage{graphicx}
\usepackage{amsmath}
\begin{document}

\title{Acoustic confinement in superlattice cavities}

\author{Daniel Garcia-Sanchez}
\email{daniel.garcia-sanchez@insp.upmc.fr}
\affiliation{Sorbonne Universit\'es, UPMC Univ.\ Paris 06, CNRS-UMR 7588, Institut des NanoSciences de Paris, F-75005, Paris, France}
\author{Samuel D\'eleglise}
\affiliation{LKB - UPMC - 4 place Jussieu, Case 74, 75252 PARIS cedex 05}
\author{Jean-Louis Thomas}
\affiliation{Sorbonne Universit\'es, UPMC Univ.\ Paris 06, CNRS-UMR 7588, Institut des NanoSciences de Paris, F-75005, Paris, France}
\author{Paola Atkinson}
\affiliation{Sorbonne Universit\'es, UPMC Univ.\ Paris 06, CNRS-UMR 7588, Institut des NanoSciences de Paris, F-75005, Paris, France}
\author{Camille Lagoin}
\affiliation{Sorbonne Universit\'es, UPMC Univ.\ Paris 06, CNRS-UMR 7588, Institut des NanoSciences de Paris, F-75005, Paris, France}
\author{Bernard Perrin}
\affiliation{Sorbonne Universit\'es, UPMC Univ.\ Paris 06, CNRS-UMR 7588, Institut des NanoSciences de Paris, F-75005, Paris, France}

\date{\today}

\begin{abstract}
The large coupling rate between the acoustic and optical fields confined in GaAs/AlAs superlattice cavities makes them appealing systems for cavity optomechanics.
We have developed a mathematical model based on the scattering matrix that allows the acoustic guided modes to be predicted in nano and micropillar superlattice cavities.
We demonstrate here that the reflection at the surface boundary considerably modifies the acoustic quality factor and leads to significant confinement at the micropillar center.
Our mathematical model also predicts unprecedented acoustic Fano resonances on nanopillars featuring small mode volumes and very high mechanical quality factors, making them attractive systems for optomechanical applications.
\end{abstract}

%\pacs{42.50.Lc, 07.10.Pz, 85.85.+j}

\maketitle

\section{Introduction}
Quantum optomechanics is a rapidly developing field of research.
The optomechanics archetypal setup consists of a high-finesse optical cavity whose optical mode is coupled to the displacement of a mechanical resonator.
In the past few years, several research groups using different device designs have reached a regime where the optical radiation pressure cools the motion of the mechanical resonator~\cite{ASchiliesserPRL2006,OArcizetNature2006,SGiganNature2006,DKlecknerNature2006}.
This technique has been used to bring the mechanical resonator to the quantum ground state~\cite{JChanNature2011,JDTeufelNature2011}.
A remarkable progress in the linear regime~\cite{MAspelmeyerRPM2014} has been achieved; highlights include optomechanically induced transparency~\cite{SWeisScience2010,AHSafaviNaeiniNature2011}, generation of squeezed light~\cite{DWCBrooksNature2012,ASafaviNaeiniNature2013,TPPurdy2013PRX}, and mechanically mediated state transfer~\cite{TAPalomakiNature2013}.
Considerable efforts are now dedicated to reaching the nonlinear regime, which holds immense promise for the study of large-scale quantum phenomena such as the preparation of nonclassical states of mechanical resonators~\cite{SBosePRA1997,QJiangPRL2012}.
In the canonical system this regime is achieved when the single photon coupling strength $g_0$ is comparable to the mechanical frequency of the resonator $\omega_m$ and the decay rate of the optical cavity $\kappa$~\cite{PRabl2011PRL,ANunnenkampPRL2011}.
These conditions are far from reach with current devices, and to enhance the nonlinearity several alternative techniques have been proposed~\cite{KStannigelPRL2012,PKomrar2013PRL,MALemondeNatureComm2016}.
These approaches still require large optomechanical coupling rates 

\begin{figure}[b]
\begin{center}
\includegraphics{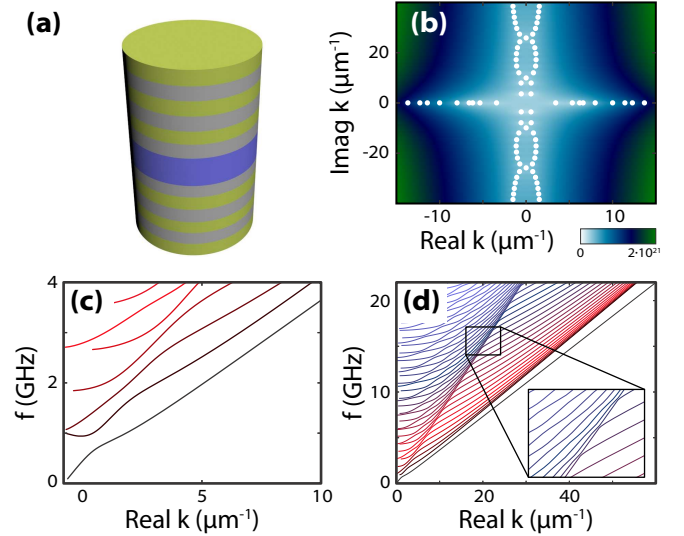}
\end{center}
\caption{\textbf{(a)} Micropillar cavity.
\textbf{(b)} Magnitude of the Pochhammer--Chree formula [given in Eq.~\eqref{eq:pochhammerchree}]
for complex wave numbers of a GaAs waveguide. $R = 1.5\,\mu\textrm{m}$  and $\omega/2\pi=5\,\textrm{GHz}$.
The roots of the Pochhammer--Chree equation are represented by white points.
\textbf{(c)} and \textbf{(d)} Dispersion curves of the real modes.}\label{fig:DBRCavityDispersionCurves}
\end{figure}

GaAs/AlAs superlattice cavities, where a $\lambda/2$ cavity is enclosed by two distributed Bragg reflector~(DBR) mirrors, can simultaneously confine the light field and the mechanical field~\cite{AFainsteinPRL2013, GRozasPRL2009}.
The great advantage of superlattice planar cavities is that the large overlap between both optical and vibrational fields in the cavity region and the photoelastic effect leads to high optomechanical coupling rates~\cite{AFainsteinPRL2013}.
Micropillar superlattice cavities can be obtained by etching a superlattice planar cavity~[see Fig.~\ref{fig:DBRCavityDispersionCurves}(a)].
Optical quality factors exceeding $200\,000$ have been demonstrated~\cite{CArnoldAPL2012,SReitzensteinAPL2007} in micropillar superlattice cavities and high quality factors for the mechanical mode are also expected and are calculated here.
In addition higher optomechanical coupling rates are expected for micropillar cavities, compared to planar superlattice cavities, because of the reduced acoustic and optical-mode volumes.
All these qualities make them excellent candidates for optomechanical experiments.

However, the acoustic confinement in micropillar cavities differs significantly from the acoustic confinement in planar cavities.
At the air-semiconductor boundary of the micropillar there is conversion between the longitudinal and transverse vibrations.
In addition the $100\,\%$ reflection which occurs at the micropillar air-semiconductor boundary surface can considerably modify the mechanical quality factor and the acoustic-mode shape.

\section{Theory}
Here we study the acoustic confinement in superlattice micropillar cavities taking into account the effect of the air-semiconductor boundary.
We have developed a semi-analytical model that combines the scattering matrix and the acoustic modal expansion of each layer of the micropillar.
First we calculate the eigenmodes of each single layer and then we impose continuity of the displacement and stress fields at the boundaries between the layers.
Each layer is modeled as an individual cylindrical acoustic waveguide and the acoustic modes are calculated along the cylinder axis $z$ in cylindrical coordinates $r$, $\theta$, and $z$.
We consider only axisymmetric modes because, for symmetry reasons, only these modes are coupled with the optical field.
Thus, the displacement and stress components are independent of $\theta$.
We can define $p=(\omega^2/c_l^2-k^2)^{1/2}$ and $q=(\omega^2/c_t^2-k^2)^{1/2}$ where $k$ is the acoustic wavenumber,  $\omega$ is the angular frequency, $c_l$ is the speed of sound for longitudinal waves, and $c_t$ is the speed of sound for transverse waves.
It can be shown~\cite{DRoyer2000} that the displacement components for the axisymmetric modes in the radial and longitudinal directions can be written in terms of the Bessel functions of the first kind:
\begin{align}
 u_r &= -\left[pA J_1(p r)+ i k C  J_1(q r)\right]e^{i (\omega t -k z)}\\
 u_z &= -\left[i k A  J_0(p r)+ q C J_0(q r)\right] e^{i (\omega t -k z)}
\end{align}
where $A$ and $C$ are related constants with $2ikpJ_1(pR)A+(q^2-k^2)J_1(qR)C=0$.
Similar expressions can be found for the stress components $\sigma_{ij}$ (see Appendix~\ref{appendix:acousticwaveguide}).

The allowed values for the wave number can be found from the dispersion relation, which is given by the Pochhammer--Chree equation~\cite{DRoyer2000}
\begin{multline}
\frac{2p}{R}(q^2+k^2)J_1(pR)J_1(qR)- (q^2-k^2)^2J_0(pR)J_1(qR)\\
-4k^2pqJ_1(pa)J_0(qR)=0\label{eq:pochhammerchree}
\end{multline}
where $R$ is the radius of the cylindrical waveguide.
The solutions of this equation can be real or complex~[see Fig.~\ref{fig:DBRCavityDispersionCurves}(b)].
The modes with a real wave number are propagative whereas the modes with a complex wave number are evanescent.
Figures~\ref{fig:DBRCavityDispersionCurves}(c) and \ref{fig:DBRCavityDispersionCurves}(d) show the dispersion curves for small and large real wavenumber ranges respectively.
The lowest frequency mode is the only one with no cut-off frequency,
therefore for small wavenumbers the waveguide is monomode with only one propagative mode and for large wavenumbers the waveguide is multimode.
For small wavenumbers the waveguide behaves as a one-dimensional solid and the speed of sound is given by $c_u=\sqrt{E/\rho}$ where $E$ is the Young's modulus and $\rho$ is the density.
For large wavenumbers the lowest frequency mode corresponds to a surface acoustic wave
and the second lowest frequency mode corresponds to a transverse mode.
None of the modes from the dispersion curve corresponds to a pure longitudinal mode, 
since a pure longitudinal mode is not an eigenmode of the system.  
However, it is possible to prepare a longitudinal wave as a linear combination of the eigenmodes that have a phase velocity close to the longitudinal phase velocity~[see inset from Fig.~\ref{fig:DBRCavityDispersionCurves}(d)].

The solutions of the Pochhammer--Chree equation of an infinite cylindrical waveguide, which we denote by $(\mathbf{\Psi}^j)=(\sigma_{\alpha\beta}^j,u_\alpha^j)$, provide an orthogonal basis of the displacement with respect to the following bilinear form:
\begin{multline}
(\mathbf{\Psi}^j,\mathbf{\Psi}^l) =\\
\int_0^R \left[\sigma_{rz}^j u_r^l-\sigma_{zz}^j u_z^l+\sigma_{rz}^l u_r^j-\sigma_{zz}^l u_z^j \right] r\, dr\label{eq:bilinearform}
\end{multline}
To demonstrate the orthogonality of the basis we calculate the relationship between the components of the modes $k_j^*$ and $k_j$.
If we set $C=1$ in Eq.~\ref{eq:relationAC} we can calculate the relations between the displacement and stress fields for $k_j^*$ and $k_j$.
\begin{align}
A^j &= \frac{(q_j^2-k_j^2)J_1(q_jR)}{2ik_jp_jJ_1(p_jR)}\\
A^{j^*} &= \frac{(q_{j^*}^2-k_{j^*}^2)J_1(q_{j^*}R)}{2ik_{j^*}p_{j^*}J_1(p_{j^*}R)}\\
A^{j^*}&=-(A^j)^*\\
 u_r^{j^*}&=-(u_r^j)^*\\
 u_z^{j^*}&=(u_z^j)^*\\
 \sigma_{rr}^{j^*}&=-(\sigma_{rr}^j)^*\\
 \sigma_{rz}^{j^*}&=(\sigma_{rz}^j)^*\\
 \sigma_{zz}^{j^*}&=-(\sigma_{zz}^j)^*
\end{align}

\begin{figure}[t]
\begin{center}
\includegraphics{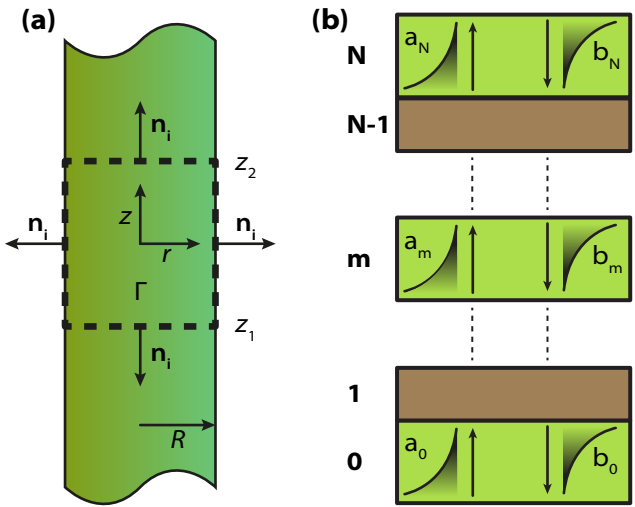}
\end{center}
\caption{\textbf{(a)} Acoustic waveguide. The volume $\Gamma$ is bounded by sections $z_1$, $z_2$  and the boundary surface of the cylinder.
\textbf{(b)} Labeling scheme for the forward and backward propagating modes $\mathbf{a}_i$ and  $\mathbf{b}_i$ in a multilayered structure.
}\label{fig:ScatteringMatrix}
\end{figure}

If there are no acoustical sources in the volume $\Gamma$ and $(\sigma_{\alpha\beta}^A,u_\alpha^{A})$ and $(\sigma_{\alpha}^B,u_\alpha^{B})$ are two solutions of the acoustics wave equation in that volume, the reciprocity theorem~\cite{IVasconcelosPRE2009} can be written in the frequency domain as
\begin{equation}
  \int_{\partial\Gamma} \left[\sigma_{\alpha\beta}^A (u_\alpha^B)^* n^\beta -(\sigma_{\alpha\beta}^B)^* u_\alpha^A n^\beta \right]dS = 0\label{eq:reciprocity_theorem}
\end{equation}
where the vector $n^\beta$ represents the outward normal to the surface $\partial\Gamma$ that encloses the volume $\Gamma$.
As shown in Fig.~\ref{fig:ScatteringMatrix}(a), we consider the cylindrical volume $\Gamma$ that is bounded by sections $z_1$, $z_2$ and the boundary surface of the cylinder.
Since we consider that the air-waveguide boundary is free, the integral is equal to zero at the waveguide boundary because
$\left.\sigma_{rz}\right|_{r=R}=\left.\sigma_{rr}\right|_{r=R}=0$ and the vector $n_i^\beta$ is perpendicular to the axis $z$.
If we apply equation \eqref{eq:reciprocity_theorem} to two solutions of the Pochhammer-Chree equation with wavenumbers $k_j$ and $k_l^*$ in the volume $\Gamma$ we obtain
\begin{multline}
  -2\pi\left[e^{-(ik_j-ik_l)z_1}-e^{-(ik_j-ik_l)z_2}\right]\times\\
  \int_0^R \left[\sigma_{rz}^j u_r^l-\sigma_{zz}^j u_z^l+\sigma_{rz}^l u_r^j-\sigma_{zz}^l u_z^j \right]rdr = 0
\end{multline}

If $k_j=k_l$ then the first factor is equal to zero, otherwise the second factor has to be equal to zero.
The bilinear form of the Eq.~\eqref{eq:bilinearform} is not a scalar product since $(\mathbf{\Psi}^j,\mathbf{\Psi}^j)$ can take negative values if $k_j$ is complex.
The bilinear form is positive definite when $k_j$ is real since in this case $(\mathbf{\Psi}^j,\mathbf{\Psi}^j)$ corresponds to the Poynting vector.
It is convenient to normalize the eigenvectors $(\mathbf{\Psi}^j)$ such that $(\mathbf{\Psi}^j,\mathbf{\Psi}^l) = \delta_{jl}$.

By using this bilinear form we can calculate the interface transfer matrices between two adjacent layers of the structure.
If we do not consider the layer thickness, the calculation of the interface matrix between layers $n$ and $n+1$ can be done by using the continuity of the displacement and stress components $\sigma_{rz}$ and $\sigma_{zz}$ at the boundary between the two layers.
In this case, it is straightforward to demonstrate that the interface matrix $\mathbf{P}(n+1)$ between the layers $n$ and $n+1$ is given by $P_{jl}(n+1)=(\Psi_n^j,\Psi_{n+1}^l)$.
If we consider that the layers have a finite thickness, the evolution matrix $\mathbf{F}(n)$ of the $n$th layer has to be taken into account.
The evolution matrix is a diagonal matrix with $F_{jj}(n)=e^{-ik_j^n\Delta z}$.
The interface matrix is given by
$\mathbf{I}(n+1)=\mathbf{F}(n)^{-1}\cdot\mathbf{P}(n+1)$.

Now we consider a multilayered waveguide with $N$ layers~[see Fig.~\ref{fig:ScatteringMatrix}(b)] and forward- and backward-propagating modes with coefficients $\mathbf{a}_i$ and  $\mathbf{b}_i$ at each interface $i$, respectively.
We use here a scattering-matrix method which provides great numerical stability~\cite{DYKoPRB1988,DMWhittaker1999}, where the coefficients of the outgoing modes $(\mathbf{a}_N,\mathbf{b}_0)$ can be related to the coefficients of the incoming modes $(\mathbf{a}_0,\mathbf{b}_N)$ via the scattering matrix $\mathbf{S}(0,N)$:
\begin{equation}
\left(
 \begin{array}{c}
  \mathbf{a}_N\\
  \mathbf{b}_0
 \end{array}
 \right)
 =
 \mathbf{S}(0,N)
 \left(
 \begin{array}{c}
  \mathbf{a}_0\\
  \mathbf{b}_N
 \end{array}
 \right)
\end{equation}

An iterative method can be used for the calculation  of the scattering matrix~\cite{DYKoPRB1988} and the acoustic fields of the layered structure~(see Appendix~\ref{appendix:scatteringmatrix}).

\section{Results}
Having established the mathematical model we can now study superlattice micropillar cavities
where the phononic mode is confined in a $\lambda/2$ cavity by two GaAs/AlAs DBR mirrors.
We define the layers such that a longitudinal resonance is expected at $20\,\textrm{GHz}$.
The thicknesses for the GaAs/AlAs pairs are $59.1\,\textrm{nm}$ and $70.4\,\textrm{nm}$ respectively.
The GaAs cavity length is $118.2\,\textrm{nm}$.

\begin{figure*}[t]
\begin{center}
\includegraphics{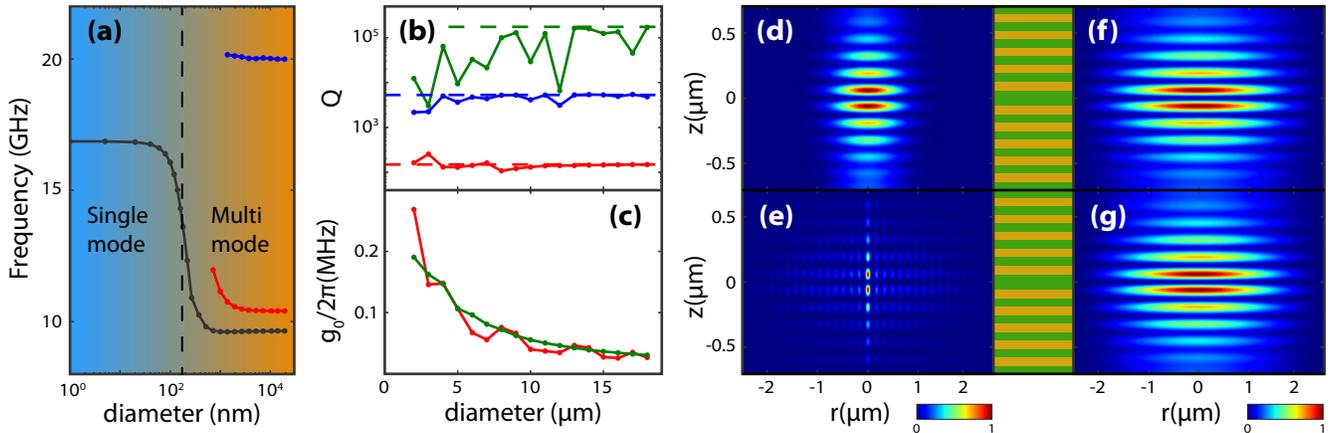}
\end{center}
\caption{\textbf{(a)} Mechanical frequency resonances of a GaAs/AlAs DBR micropillar cavity as a function of diameter:
The fundamental mode (black), transverse mode (red), and longitudinal mode (blue).
\textbf{(b)} Quality factor of the longitudinal mode as a function of diameter.
Ten pairs per mirror (red), 20 pairs per mirror (blue) and 30 pairs per mirror (green).
Dashed lines represent the quality factor in a planar structure.
\textbf{(c)} Coupling rate $g_0$ of the longitudinal mode as a function of the diameter.
Ten pairs per mirror (red) and 30 pairs per mirror (green).
\textbf{(d)} Squared displacement of the acoustic longitudinal mode for a micropillar with a diameter of $5\,\mu\textrm{m}$ and five pairs per mirror.
\textbf{(e)} Squared displacement of the acoustic longitudinal mode. 25 pairs per mirror.
\textbf{(f)} Squared electrical field of the optical mode. 5 pairs per mirror.
\textbf{(g)} Squared electrical field of the optical mode. 25 pairs per mirror.}\label{fig:DBRCavity}
\end{figure*}

All the modes with a real wave number are considered in the calculation, and evanescent modes are included to ensure  continuity of the displacement and stress at the layer boundaries. A numerical cut off is applied for the high-wave-number evanescent modes once their  contribution becomes negligible.
The calculations for a $15\,\mu\textrm{m}$ diameter micropillar requires 157 real wavenumber modes and 920 complex wave-number modes.
If the diameter is smaller than the wavelength, the micropillar is single mode.
In the limit of a very small diameter the speed of sound is given by $c_u$.
A one-dimensional transfer-matrix calculation using $c_u$ shows that the mechanical resonance should be found at $17\,\textrm{GHz}$, which coincides with the value obtained with our mathematical model~[see Fig.~\ref{fig:DBRCavity}(a)]. 
If the diameter is bigger than the wavelength, the micropillar is multimode: in our structure the pillar is multimode for diameters bigger than $150\,\textrm{nm}$.
The fundamental mode at $9.6\,\textrm{GHz}$ for the larger diameter pillars
corresponds to the surface-acoustic-wave resonance and the mode at $10.4\,\textrm{GHz}$ corresponds to the transverse-mode resonance.
These two resonance modes correspond well with individual modes in the dispersion curve from Fig.~\ref{fig:DBRCavityDispersionCurves}(c),
indicating that only a single eigenmode per layer participates in the micropillar resonance.
This is not the case for the longitudinal mode that can be found at $20\,\textrm{GHz}$ which, as discussed earlier, is formed by a combination of the modes that have a phase velocity close to the longitudinal phase velocity~[see inset from Fig.~\ref{fig:DBRCavityDispersionCurves}(d)].
To calculate the resonance frequency of the longitudinal mode we excite the first layer with a longitudinal Gaussian beam that has been constructed as a linear combination of the eigenmodes given by the solution of the Pochhammer--Chree equation.
The calculation of the longitudinal and transverse resonance modes using a one-dimensional transfer-matrix calculation~\cite{AFainsteinPRL2013} gives the same result as our model in the limit of a large diameter.

Now we do an in-depth analysis of the longitudinal mode.
This mode provides the highest coupling rate with the optical field because longitudinal vibrations modify considerably the effective length of the optical cavity whereas the contribution of transverse vibrations to the optomechanical coupling rate is minimum.
The air-semiconductor boundary of the micropillar, which is a perfect reflector, can modify the shape and the mechanical quality factor of the longitudinal mode depending on the reflectivity of the upper and lower DBR mirrors, as shown in Figs.~\ref{fig:DBRCavity}(d) and \ref{fig:DBRCavity}(e).
We can define $l_c=Q\lambda$ as the length traveled by vibrations inside the cavity %\textbf{(check that this is correct!. ex factor 2, or pi)}
and $z_R$ as the Rayleigh length of the acoustic Gaussian beam.
If the number of pairs is small, the mechanical quality factor is small, $z_R$ is bigger than $l_c$, and the resonance mode has a Gaussian shape because there is no reflection at the boundaries [see~Fig.~\ref{fig:DBRCavity}(d)].
In this case the micropillar can be modeled as a one-dimensional device.
As shown in Fig.~\ref{fig:DBRCavity}(b), in this regime the quality factor obtained with a one-dimensional transfer matrix calculation~\cite{AFainsteinPRL2013} coincides with the quality factor obtained with the mathematical model from this paper; this validates our model.
If the number of pairs per mirror is increased, the quality factor is increased and $l_c$ becomes much larger than $z_R$;
as a result the mode volume is considerably reduced because of the reflections at the air-semiconductor boundary of the micropillar [see Figure~\ref{fig:DBRCavity}(e)].

As shown in Fig.~\ref{fig:DBRCavity}(b) the reflections at the air-semiconductor boundary can considerably modify the quality factor.
Figure~\ref{fig:DBRCavity}(b) shows that the mechanical quality factor can be improved by increasing the number of pairs per mirror, but this improvement is considerably reduced for small micropillar diameters.
The coupling rate $G$, representing the optical resonance frequency shift per unit displacement amplitude, is reduced with the number of pairs per mirror because the overlap between the mechanical mode from Figs.~\ref{fig:DBRCavity}(d) and \ref{fig:DBRCavity}(e) with the optical mode from Figs.~\ref{fig:DBRCavity}(f) and \ref{fig:DBRCavity}(g) is significantly reduced.
For the calculation of the optical field we have used a semi-analytical approach~\cite{YLDHoIEEEJQE2007}.
On the other hand the zero-point amplitude of the mechanical resonator, $x_\mathrm{zpf}$, increases with the number of pairs per mirror because the mode volume and the effective mass decreases.
Interestingly, as shown in Fig.~\ref{fig:DBRCavity}(c), these two effects compensate each other and $g_0=G\cdot x_\mathrm{zpf}$ is almost independent of the number of pairs per mirror.
The optomechanical coupling rate $g_0$ has been calculated with a perturbative approach~\cite{MEichenfieldNature2009,JChanAPL2012,SJohnsonPRE2002} which accounts for the geometric and the photo-elastic contributions.
We used bulk photoelastic coefficients for GaAs, $(p_{11},p_{12},p_{44})=(-0.165, 0.140, -0.072)$~\cite{CBakerOE2014}.
With these values, the photoelastic contribution to $g_0$ is roughly $15\%$.
However, large photoelastic coefficients have been observed in GaAs/AlAs multiple quantum wells close to the exciton-polariton resonance~\cite{BJusserand2015PRL}.

\begin{figure}[t]
\begin{center}
\includegraphics{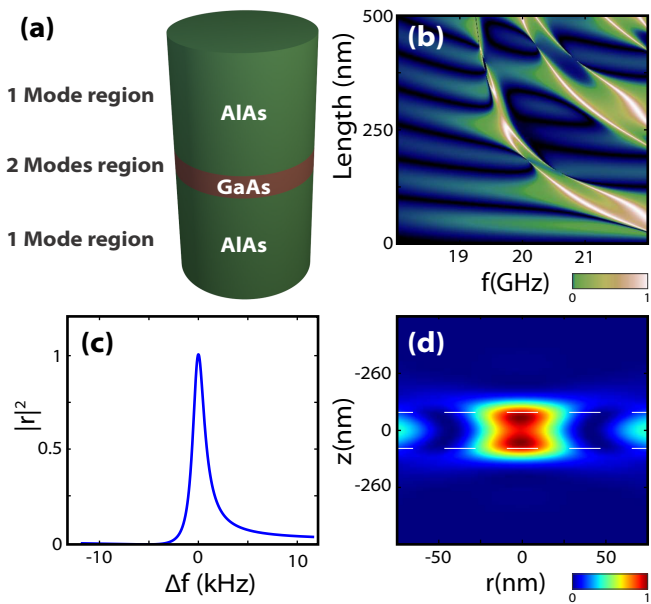}
\end{center}
\caption{\textbf{(a)} Schematic of a Fano cavity. A bimodal layer is enclosed by two infinite single-mode waveguides.
The diameter is fixed at $150\,\textrm{nm}$.
The central layer material is GaAs and the upper and lower cladding layers are AlAs.
\textbf{(b)} Acoustic reflectivity of the Fano cavity as a function of the frequency and the cavity length.
\textbf{(c)} Reflectivity vs deviation from the resonant frequency showing a Fano lineshape. %%Fano resonance.% 
The cavity length is $182\,\textrm{nm}$ and the resonance quality factor is $1.3\cdot 10^6$.
\textbf{(d)} Displacement mode shape of the resonance.
The lines represent the interface between the regions.}\label{fig:FanoRegime}
\end{figure}

To obtain a small mechanical mode volume and a large $g_0$, the micropillar diameter has to be considerably reduced,
but as shown above, it is not possible to obtain high mechanical quality factors with small micropillar diameters.
Here we propose an alternative approach that allows high mechanical quality factors with very small mode volumes to be obtained.
The structure consists of a GaAs region with two propagative modes that is enclosed by two monomode AlAs regions [see Figure~\ref{fig:FanoRegime}(a)].
The first propagative mode of the GaAs region is very well coupled to the propagative mode of the AlAs region because the spatial mode shapes are very well matched.
The impedance mismatch at the GaAs/AlAs boundary leads to small reflections,
resulting in smooth oscillations of the transmission as a function of the GaAs region thickness [see Figure~\ref{fig:FanoRegime}(b)].
On the other hand the second propagative mode of the GaAs region is not well coupled to the single propagative mode of the AlAs region and is strongly reflected at the interface.
As a result the second propagative mode of the GaAs region is well confined, giving rise to high-\emph(Q) trapped modes.

The result of the interaction between these two 
transmission pathways from GaAs to AlAs leads to Fano lineshape resonances~\cite{UFanoPRL1961} characterized by an asymmetrical lineshape that has been observed in various physical systems ranging from carbon nanotubes~\cite{JKim2003PRL} to intersubband transitions in coupled quantum well systems~\cite{JFaistOL1996}.
Fig.~\ref{fig:FanoRegime}(c) shows a Fano resonance from a $182\,\mathrm{nm}$ GaAs layer embedded between AlAs cladding layers. The resonance has a quality factor of $1.3\times 10^6$ and also has a very small mode volume as can be seen in Fig.~\ref{fig:FanoRegime}(d).
In this device there is no optical cavity, but it is possible to couple the mechanical mode through strain~\cite{IYeo2014NatNano,JTeissierPRL2015} to the optical transition of a quantum dot that can be placed in the GaAs region.
This would allow the mechanical mode to be cooled down to the ground state~\cite{IWilsonRaePRL2004, MMetcalfePRL2010} and has interesting applications such as quantum state transfer between the mechanical resonator and the two level system~\cite{PRablNatPhys2010}.

\section{Conclusion}
The mathematical model presented here provides the foundation for the acoustic analysis of layered structures with lateral confinement.
Our calculations show that the total reflection of vibrational modes at the air-semiconductor boundaries of superlattice micropillar cavities has a dramatic effect on the acoustical quality factor and mode volume. In particular, the acoustic-mode volume is strongly dependent on the number of mirror repeats, unlike the optical-mode volume.
Our model reveals as well novel acoustic Fano cavities based on a simple waveguide structure patterned into nanopillars,
which are an attractive alternative to superlattice micropillar cavities for optomechanical experiments.

\section{Acknowledgments}
This work was supported supported by the ``Agence Nationale de la Recherche'' and the ``Idex Sorbonne Universit\'es'' under contract No.\ ANR-11-IDEX-0004-02, through the MATISSE program.

\appendix
\section{Cylindrical acoustic waveguide}\label{appendix:acousticwaveguide}
The displacement vector $\mathbf{U}$ of a cylindrical acoustic waveguide can be derived from a scalar potential $\phi$ and a vector potential $\Psi$~\cite{DRoyer2000}.
These two potentials satisfy the following wave equations
\begin{equation}
 \nabla^2\phi -\frac{1}{c_l^2}\frac{\partial^2 \phi}{\partial t^2}= 0 \text{\hspace{0.2cm} and \hspace{0.2cm}} \nabla^2\Phi -\frac{1}{c_t^2}\frac{\partial^2 \Phi}{\partial t^2}= 0 
\end{equation}
where $c_l$ and $c_t$ are the speed of sound for bulk longitudinal and transverse waves.

It is convenient to adopt the cylindrical coordinates $r$, $\theta$ and $z$.
We consider only the axisymmetric modes because, for symmetry reasons, only these modes are coupled with the optical field.
The displacement and stress components of the modes are independent of $\theta$ and can be written as
\begin{align}
 U_\alpha(r,z,t) &= \frac{1}{2\pi}\int_{\omega} u_\alpha e^{i (\omega t -k z)}d\omega\\
 \Sigma_{\alpha\beta}(r,z,t) &= \frac{1}{2\pi}\int_{\omega} \sigma_{\alpha\beta}  e^{i (\omega t -k z)}d\omega
\end{align}
where $k$ represents the wave number, $\omega$ is the angular frequency, and $t$ is the time.
The coefficients of the Fourier transform are given by~\cite{DRoyer2000}
\begin{align}
u_r &= -\left[pA J_1(p r)+ i k C  J_1(q r)\right]\\
u_z  &=-\left[i k A  J_0(p r)+ q C J_0(q r)\right]\\
\sigma_{rr} &=  -AJ_0(p r) (\lambda  k^2  +\lambda  p^2  +2 \mu  p^2)\\
   &\phantom{{}={}} +\frac{2\mu  p AJ_1(p r)-2 i C k \mu  (q r J_0(q r)+J_1(q r)) }{r}\nonumber\\
\sigma_{rz} &=\mu  \left[2 i A k p J_1(p r) + C \left(q^2-k^2\right) J_1(q r)\right] \\
\sigma_{zz} &= [A J_0(p r)\left(-\lambda  k^2 -2 k^2 \mu  -\lambda  p^2 \right)\nonumber\\
&\phantom{{}={}}+2 i C k \mu  q J_0(q r)]
\end{align}
where $p=(\omega^2/c_l^2-k^2)^(1/2)$ and $q=(\omega^2/c_t^2-k^2)^(1/2)$.

At the boundary surface of the cylindrical waveguide, the stress components $\sigma_{rr}$ and $\sigma_{rz}$ are equal to zero:
\begin{align}
0 &= \left[-(q^2-k^2)J_0(pR)+2\frac{p}{R}J_1(pR)\right]A\nonumber\\
&\phantom{{}={}}+2ik\left[-qJ_0(qR)+\frac{1}{R}J_1(qR)\right]C\\
0 &=  2ikpJ_1(pR)A+(q^2-k^2)J_1(qR)C\label{eq:relationAC}
\end{align}
where $R$ represents the radius of the waveguide.
These conditions are satisfied only if the determinant is equal to zero, giving the dispersion relation for the wave vector, also known as the Pochhammer--Chree equation:
\begin{multline}
\frac{2p}{R}(q^2+k^2)J_1(pR)J_1(qR)- (q^2-k^2)^2J_0(pR)J_1(qR)\\
-4k^2pqJ_1(pa)J_0(qR)=0
\end{multline}

\section{Scattering matrix}\label{appendix:scatteringmatrix}
To avoid numerical instabilities the scattering matrix approach is used~\cite{DYKoPRB1988,DMWhittaker1999}.
For a \emph{N}-layer structure, the coefficients $\mathbf{a}_N$ and $\mathbf{b}_0$ of the outgoing modes are related to the coefficients $\mathbf{a}_0$ and $\mathbf{b}_N$ of the incoming modes via the scattering matrix $\mathbf{S}(0,N)$
\begin{equation}
\left(
 \begin{array}{c}
  \mathbf{a}_N\\
  \mathbf{b}_0
 \end{array}
 \right)
 =
 \mathbf{S}(0,N)
 \left(
 \begin{array}{c}
  \mathbf{a}_0\\
  \mathbf{b}_N
 \end{array}
 \right)
\end{equation}

The matching conditions for the wave functions at the (n + 1)th interface can be expressed as function of the interface matrix:
\begin{equation}
\left(
 \begin{array}{c}
  \mathbf{a}_n\\
  \mathbf{b}_n
 \end{array}
 \right)
 =
 \mathbf{I}(n+1)
 \left(
 \begin{array}{c}
  \mathbf{a}_{n+1}\\
  \mathbf{b}_{n+1}
 \end{array}
 \right)
\end{equation}

It is convenient to divide the scattering and interface matrices in submatrices
\begin{align}
\mathbf{S}
 &=
 \left(
 \begin{array}{cc}
  \mathbf{S}_{11}\mathbf{S}_{12}\\
  \mathbf{S}_{21}\mathbf{S}_{22}
 \end{array}
 \right)\\
 \mathbf{I}
 &=
 \left(
 \begin{array}{cc}
  \mathbf{I}_{11}\mathbf{I}_{12}\\
  \mathbf{I}_{21}\mathbf{I}_{22}
 \end{array}
 \right)
\end{align}

To calculate the transfer matrix $\mathbf{S}(0,N)$ an iterative process based on the following expressions~\cite{DYKoPRB1988,DMWhittaker1999} can be used:
\begin{align}
 \mathbf{S}_{11}(0,n+1)&=(\mathbf{I}_{11}-\mathbf{S}_{12}\mathbf{I}_{21})^{-1}\mathbf{S}_{11}\\
 \mathbf{S}_{12}(0,n+1)&=(\mathbf{I}_{11}-\mathbf{S}_{12}\mathbf{I}_{21})^{-1}(\mathbf{S}_{12}\mathbf{I}_{22}-\mathbf{I}_{12})\\
 \mathbf{S}_{21}(0,n+1)&=\mathbf{S}_{22}\mathbf{I}_{21}\mathbf{S}_{11}(0,n+1)+\mathbf{S}_{21}\\
 \mathbf{S}_{22}(0,n+1)&=\mathbf{S}_{22}\mathbf{I}_{21}\mathbf{S}_{12}(0,n+1)+\mathbf{S}_{22}\mathbf{I}_{22}
\end{align}

To calculate the stress and displacement fields of the whole structure the coefficients of all the layers have to be calculated by using the expressions~\cite{DYKoPRB1988,DMWhittaker1999}
\begin{align}
 \mathbf{a}_n&=(1-\mathbf{S}_{12}(0,n)\mathbf{S}_{21}(n,N))^{-1}\nonumber\\
 &\phantom{{}={}}(\mathbf{S}_{11}(0,n)\mathbf{a}_0+\mathbf{S}_{12}(0,n)\mathbf{S}_{22}(n,N)\mathbf{b}_N)\\
 \mathbf{b}_n&=(1-\mathbf{S}_{21}(n,N)\mathbf{S}_{12}(0,n))^{-1}\nonumber\\
 &\phantom{{}={}}(\mathbf{S}_{21}(n,N)\mathbf{S}_{11}(0,n)\mathbf{a}_0+\mathbf{S}_{22}(n,N)\mathbf{b}_N)
\end{align}

\end{document}